\documentclass[10pt, conference]{IEEEtran}
\usepackage[utf8]{inputenc}
\usepackage{verbatim}
\usepackage{listings}
\usepackage{braket}
\usepackage{enumitem, kantlipsum}
\usepackage{graphicx}
\usepackage[dvipsnames,table,xcdraw]{xcolor}
\usepackage{multicol}
\usepackage{mathtools}

\usepackage{csquotes}
\usepackage{amsmath}

\usepackage{url}

\usepackage{qcircuit}
\usepackage{amsthm,amssymb}
\usepackage{fdsymbol}
\usepackage{rotating}
\usepackage[symbol]{footmisc}

\usepackage{arydshln}

\let\OLDthebibliography\thebibliography
\renewcommand\thebibliography[1]{
  \OLDthebibliography{#1}
  \setlength{\parskip}{0pt}
}

\title{\huge ACSS-q: Algorithmic complexity for short strings\\via quantum accelerated approach}
\usepackage[affil-it]{authblk}

\author[1]{Aritra Sarkar}
\author[2]{Koen Bertels}
\affil[1]{Department of Quantum \& Computer Engineering,
Faculty of Electrical Engineering, Mathematics and Computer Science,
\protect\\Delft University of Technology, Delft, The Netherlands}
\affil[2]{Department of Informatics Engineering,
Faculty of Engineering,
\protect\\University of Porto, Porto, Portugal}
\date{} %\today}

\begin{document}

\maketitle

\begin{abstract}
    % \footnote[1]{Unpublished work in progress, DO NOT DISTRIBUTE}
    In this research we present a quantum circuit for estimating algorithmic complexity using the coding theorem method.
    This accelerates inferring algorithmic structure in data for discovering causal generative models.
    The computation model is restricted in time and space resources to make it computable in approximating the target metrics.
    The quantum circuit design based on \cite{sarkar2020quantum} that allows executing a superposition of automata is presented.
    As a use-case, an application framework for protein-protein interaction ontology based on algorithmic complexity is proposed.
    Using small-scale quantum computers, this has the potential to enhance the results of classical block decomposition method towards bridging the causal gap in entropy based methods.
    % Our implementation on the OpenQL quantum programming language and the QX Simulator is copy-left and can be found on \url{https://github.com/Advanced-Research-Centre/QuBio}
    
\end{abstract}


\section{Introduction} \label{sec:introduction}

The ubiquitous nature of algorithmic metrics is juxtaposed with it uncomputablility.
Algorithmic (Kolmogorov) complexity, algorithmic (Solomonoff) probability, universal prior (Solomonoff) distribution, universal (Levin) search, logical (Bennett) depth, omega (Chaitin) constant, etc. are all uncomputable using universal computing models, like Turing machines, or cellular automata.
Thus, while in the field of theoretical computer science, these fundamental concepts are valuable for proofs, their applicability to real-world data and use-cases remain very limited.

For example, we find forms of universal search and universal prior being used in artificial general intelligence models like the Jurgen Schmidhuber's Godel Machines and Marcus Hutter's AIXI.
While there has been attempts in making AIXI computable using approximating like AIXI-tl and UCAI, their use on real data still require enormous computing power.
This computation gap is also now being realized in data-driven models alike~\cite{thompson2020computational} and it might be worth exploring what radically different technologies like quantum computing can offer.
It is crucial that we realize all the algorithmic metrics are uncomputable, thus much higher in the computing hierarchy that the more familiar polynomial-time (P) and non-deterministic polynomial time (NP) complexity classes.
Since quantum computers are not expected to provide exponential speedup for NP-hard problems, we expect at best a quantum (Grover) search type polynomial speedup.

Often polynomial speedup between different computing models of the same power are heavily influenced by the formulation of a problem and the model itself.
For example, it is highly non-trivial to do a binary search on a Turing machine without a random-access-memory.
So, in this work we explore the possibility of a quantum algorithm formulation that retains the speedup.
Secondly, for real-world data, often a polynomial speedup is enough in making an application tractable.
In the era of small-scale quantum computers, it is thus important to explore polynomial speedups as well and better understand average case complexity and other linear and constant factors in the complexity.
In the field of quantum computing we found very few results for algorithmic metrics.
A recent research formulates a AIXI-q (quantum accelerated) for reinforcement learning, however it doesn't provide a detailed gate (time) and qubit (space) level analysis.

Our focus in this research is on accelerating the estimation of algorithmic (Kolmogorov) complexity.
This is an absolute measure of the information content of a string, defined as the shortest program for an universal Turing machine that generates that string on the output tape on halting.
For the generic case, it is uncomputable due to the halting problem.
The measure is also dependent of the exact definition of the automaton but is within a constant factor due to the invariance theorem adding the cross-compiler size.
A seminal result~\cite{Delahaye2007OnTK} in this direction considers Experimental Algorithmic Information Theory (EAIT) based on the property that many of the algorithmic metrics are semi-computable and converges in the limit.
This is crucial as otherwise any hope of approximating these quantities with computing models would not be possible.
An important distinction to point out here is that, in this research we present  'quantum accelerated estimation of the algorithmic complexity of classical sequences'.
This is different from 'quantum Kolmogorov complexity' which deals with the metrics to define the complexity of a specific quantum state and its preparation process.

There has been very few attempts to compute the uncomputable algorithmic metrics.
These attempts approximately estimation the Chaitin Omega number [4], Bennett’s Logical Depth [23,27], Universal Distribution [my paper].
Estimation of the algorithmic (Kolmogorov) complexity of sequences is actively researched by Hector Zenil and been applied to evaluations of structures in DNA, psychometrics [15], cellular automata [26,33], graph theory, economic time series [20,34] and complex networks[15].
Thus we find it promising to develop a quantum computing approach, as any benefit will be magnified by the wide applicabilty of this technique.

\cite{soler2014calculating} uses the Coding Theorem Method to estimate the algorithmic complexity of short strings.
This will be discussed in further detail in the Section~\ref{sec:compmodel}, as this forms the classical kernel that we attempt to accelerate with a quantum formulation.
In Section~\ref{sec:implementation} the quantum implementation called ACSS-q is presented.
This is based on the algorithm defined rigourously in \cite{sarkar2020quantum} for enumerating Turing machine runs in superposition.
The infrastructure required for implement a specialized quantum accelerator~\cite{bertels2020quantum} for this algorithm is discussed.
We discuss some use-cases in Section~\ref{sec:applications} motivated by our research in genome sequencing\cite{sarkar2019algorithm,sarkar2020quaser} and analysis.
In Section\ref{sec:conclusion} we conclude the paper. % Introduction
% \newpage
\section{Computation principle} \label{sec:compmodel}

\subsection{Block Decomposition Method}

Estimating the complexity of real-world data needs a scalable approach for long data streams or graphs.
To estimate the complexity of long strings and large adjacency matrices, a method based upon algorithmic probability, called Block Decomposition Method (BDM) is generally used. 
For example, the online tool (developed by Algorithmic Nature Group and the Algorithmic Dynamics Lab) called the Online Algorithmic Complexity Calculator (OACC) used BDM to provide estimations of algorithmic complexity and algorithmic probability for short and long strings and for 2-dimensional arrays, and of logical depth for strings.

BDM approximates the algorithmic complexity value for an arbitrarily large object by decomposing into smaller slices of appropriate sizes for which the values are known and aggregated back to a global estimate.
The algorithmic complexity of these small slices are precomputed and make into a look-up table using the Coding Theorem Method (CTM).
$$BDM(s) = \sum_{s_i} AC(s_i)$$
where, $AC(s_i)$ is the algorithmic complexity of the string $s_i$ and 
$$s = \cup_i s_i$$
i.e., the $s_i$ together forms the string $s$.
Small variations on the method of dividing the string into blocks becomes negligible in the limit, e.g. to take a sliding window or blocks.

Techniques based on algorithmic metrics can detect causal gaps in entropy based methods (e.g. Shannon entropy or compression algorithms).
BDM is a smooth transition between the Kolmogorov entropy and Shannon entropy, depends respective on whether the size of $s_i$ is same as $s$ or $1$.
Calculating the CTM value gets exponentially difficult with string size, thus the BDM is most likely be the method of choice for using algorithmic metrics for causal inferences.
However, we see for longer data like real-world use-cases, BDM starts getting less effective (for a fixed block size) and fails to find causal links beyond the block size.
To maintain the advantage over the simpler data-driven approaches we need to have an application motivated block size, e.g. how far in a genome can one gene affect the other.
This motivates our research in using quantum-acceleration to extend the lookup table of the Algorithmic Complexity of Short Strings (ACSS)~\cite{acssr2020}.
The quantum technique is not to supplant the BDM but to power it with the resources necessary from the quantum-accelerated CTM.

\subsection{Coding Theorem Method}

Here we will discuss the CTM and how it is used to estimate the algorithmic complexity.
The algorithmic complexity (AC), also called Kolmogorov-Chaitin complexity, is the length of the shortest program that produces a particular output and halts on a specific TM.
Formally, it is defined as:
$$AC_T(s) = \text{min:}\{|p|,T(p)\rightarrow s\}$$
where $T$ is a TM, $p$ is a program, $|p|$ is the length of the program, $s$ is a string $T(p)\rightarrow s$ denotes the fact that $p$ executed $T$ outputs $s$ and halts.
    
Algorithmic complexity is not a computable quantity in the general case due to fundamental limits of computations that arise from the halting problem (impossibility to determine whether any given program will ever halt without actually running this program, possibly for infinite time).
However, it has bounded lower-approximate property, i.e. if we can find a program $p_l$ in a language $l$ (e.g. Java, Assembly), $|p_l| \ge |p|$.
    
Since AC depends on the particular model of computation (i.e. the TM and the language), it is always possible to design a language where a particular string $s$ will have a short encoding no matter how random.
However, the invariance theorem guarantees that, there exists a constant additive factor $c_{T1\rightarrow T2}$ independent of $s$, such that
$$AC_{T2} \le AC_{T1} + c_{T1\rightarrow T2}$$
This constant is the compiler length for translating any arbitrary program $p_{T1}$ for $T1$ to $p_{T2}$ for $T2$.

The CTM used algorithmic probability to estimate the complexity as these notions are inherently interlinked.
Algorithmic probability (AP) is the chance that a randomly selected program will output $s$ when executed on $T$. 
The probability of choosing such a program is inversely proportional to the length of the program.
$$AP_T(s)=\sum_{p:T(p)\rightarrow s} 2^{-|p|}$$
Thus, the largest contribution to the term comes from the shortest program that satisfies the condition.
It is uncomputable for the same reasons as AC, and is related to AC via the following law:
$$AC_T(s)=-log_2(AP_T(s))+c$$
i.e., if there are many long programs that generate a dataset, then there has to be also a shorter one.
The arbitrary constant is dependent on the choice of a programming language.
    
AP can be approximated by enumerating TM of a given type and count how many of them produce a given output and then divide by the total number of machines that halt.
When exploring machines with $n$ symbols and $m$ states algorithmic probability of a string $s$ can be approximated as:
$$D(n,m)(s)=\dfrac{|{T\in(n,m):T \text{ outputs } s}|}{|{T\in(n,m):T \text{ halts } }|}$$
        
The Coding Theorem Method (CTM) approximates the AC  using the approximation of the AP as:
$$CTM(n,m)(s)=-log_2D(n,m)(s)$$
Calculating CTM, although theoretically computable, is extremely expensive in terms of computation time.
The space of possible Turing machines may span thousands of billions of instances.

The ACSS database used in the OACC for BDM is built using this CTM technique.
This database is constructed by approximating the output frequency distribution for Turing machines with $5$ states and $2$ symbols generating the algorithmic complexity of strings of size $\le 12$ over $\approx 500$ iterations.
These numbers of $5$, $2$, $12$ and $500$ are arbitrary from the BDM application perspective and fixed based on the computational resource.
Quoting the paper "The calculation presented herein will remain the best possible estimation for a measure of a similar nature with the technology available to date, as an exponential increase of computing resources will improve the length and number of strings produced only linearly if the same standard formalism of Turing machines used is followed."
This immediately translates to the promise of mapping this exponentially growing states using quantum computation for the Turing machine enumeration.
Thus a quantum-accelerated version on larger strings can be integrated and used immediately within the application pipeline.

The symbol set is taken as the binary alphabet, used in data encoding in current digital computation due to its simplicity.

The argument for $5$ state is more directly related to computability restrictions.
Being uncomputable, every Turing machine 'in-principle' needs to be run forever to know if it will halt.
However, for small Turing machines, it is possible to enumerate and classify every Turing machine as either it halts with a maximum time step of $B$, or goes on forever.
This is called the Busy Beaver runtime for the particular specification.
This is used in CTM to judiciously stop the computation and declare the enumeration as non-halting if it hasn't halted in $B$ steps.
The value of $B$ is known to be $107$ for Turing machines with 4 states and 2 symbols.
However, for 5 states the value is still unknown as there are too many cases to enumerate ($26559922791424$) and from partial enumerations so far, we know value is $\ge 47176870$.
It is intractable to run so many machines for so long iterations.
The paper tries to estimate the AC when the $B$ value is unknown by estimating for $500$ steps.
This is based on \cite{calude2006most} and hypotheses an exponential decay in the number of halting TM with steps with the random variable model of $P(S=k|S\le S_{lim})=\alpha e^{-\lambda k}$.
For the 5,2 case with $S_{lim}=500$ is $6*10^{-173}$, thus can be safely ignored.

The $500$ run limit considered is an informed guess based on the ability to capture on the output tape almost all $12$ bit strings (except 2) thus allowing the calculation of their CTM values.
However for $13$ bits strings, only half of all possible strings were there, setting the block size limit using this method.

The technique used various intelligent optimizations like symmetries, look ahead and patterns.
Only 4/11 of all machines were required to be enumerated for $500$ steps (if they didn't halt before), which still took $18$ days using $25*84-64$ CPUs running at $2128$ MHz with $4$ GB of memory each (on a supercomputer at the Centro Informático Científico de Andalucía (CICA), Spain).

% Computation model
% \newpage
\section{Quantum implementation} \label{sec:implementation}

\subsection{Automata model}

Rado's~\cite{rado1962non} Busy Beaver game considers $n$-cards generating $(4(n+1))^{2n}$ Turing machines.
This is slightly different when considering the number of states $m$ as the variable, giving $(4m+1)^{2m}$ binary 1-dimensional TM.
The ACSS using CTM is based on this generic Turing machine model.
In our model, we do not consider a special halt state and consider encoding the states and symbols in binary.
The number of programs is given by the equation
$$P = 2^{(m*n)*(\left \lceil{log_2(m)}\right \rceil +\left \lceil{log_2(n)}\right \rceil+d)}$$

The canonical binary alphabet $\Gamma: \{0,1\}$, with, $n = 2$ and $d = 1$ tape dimension is considered.
For $d = 1$ and $n = 2$
$$P = 2^{(2m)*(\left \lceil{log_2(m)}\right \rceil+2)} = 4^{m\left \lceil{log_2(m)}\right \rceil}*4^{2m}$$
\begin{table}[ht]
    \caption{First few values of P}
    \centering
    \begin{tabular}{|l|l|}
        \hline
        m states & number of programs P \\
        \hline
        1 & 16\\
        2 & 4096\\
        3 & 16777216\\
        4 & 4294967296\\
        5 & 1125899906842624\\
        6 & 1152921504606846976\\
        7 & 1180591620717411303424\\
        8 & 1208925819614629174706176\\
        9 & 324518553658426726783156020576256\\
        \hline
    \end{tabular}
    \label{tab:numP}
\end{table}

Note that, the state and symbols being encoded in binary, thus the number of TM is more (for states that are not exact powers of 2) than the standard value which doesn't consider how to represent the transition function on a computer.
This means there will be spurious machines in our encoding using bits/qubits where we transition from/to states that is invalid, e.g. 101, 110, 111 for a 5-state TM.

As a simple case, let the number of states $m = 1$ with the state set $Q: \{Q_0\}$.
The 16 programs for a 2 symbol TM using this description number encoding is shown on Table~\ref{tab:utm121}.

\begin{table}[ht]
    \caption{Case 1-2-1 exhaustive UTMs}
    \centering
    \begin{tabular}{|c|c:c|c}
        \hline
        P\# & $Q_0 R_1$ & $Q_0 R_0$ \\
        \hline
        00 & $Q_0 M_l W_0$ & $Q_0 M_l W_0$ \\
        01 & $Q_0 M_l W_0$ & $Q_0 M_l W_1$ \\
        02 & $Q_0 M_l W_0$ & $Q_0 M_r W_0$ \\
        03 & $Q_0 M_l W_0$ & $Q_0 M_r W_1$ \\
        \hdashline
		04 & $Q_0 M_l W_1$ & $Q_0 M_l W_0$ \\
        05 & $Q_0 M_l W_1$ & $Q_0 M_l W_1$ \\
        06 & $Q_0 M_l W_1$ & $Q_0 M_r W_0$ \\
        07 & $Q_0 M_l W_1$ & $Q_0 M_r W_1$ \\
        \hdashline
		08 & $Q_0 M_r W_0$ & $Q_0 M_l W_0$ \\
        09 & $Q_0 M_r W_0$ & $Q_0 M_l W_1$ \\
        10 & $Q_0 M_r W_0$ & $Q_0 M_r W_0$ \\
        11 & $Q_0 M_r W_0$ & $Q_0 M_r W_1$ \\
        \hdashline
		12 & $Q_0 M_r W_1$ & $Q_0 M_l W_0$ \\
        13 & $Q_0 M_r W_1$ & $Q_0 M_l W_1$ \\
        14 & $Q_0 M_r W_1$ & $Q_0 M_r W_0$ \\
        15 & $Q_0 M_r W_1$ & $Q_0 M_r W_1$ \\
        \hline
    \end{tabular}
    \label{tab:utm121}
\end{table}

\subsection{Quantum formulation}

As a computing system, Turing machine forms the theoretical model.
The input for the computation is fed in as the initial tape configuration.
The processing/program is defined by a finite state machine encoded as a state-transition table based on which the Turing machine reads the tape, changes it's state based on it's current state and the input, writes back on the tape, moves the tape head and updates its state.
The final tape when the machine halts is the output.

An example of a 2-symbol Turing machine that adds 2 unary (sequence of 1s) numbers separated by 0, using 4 states is as shown in Table~\ref{tab:tmadd}.

\begin{table}[ht]
    \caption{TM transition table to add two unary numbers}
    \centering
    \begin{tabular}{|c|c|}
        \hline
        $Q_t R_t$ & $Q_{t+1} M_t W_t$ \\
        \hline
        $Q_0 R_0$ & $Q_1 M_r W_1$ \\
        $Q_0 R_1$ & $Q_0 M_r W_1$ \\
        $Q_1 R_0$ & $Q_2 M_l W_0$ \\
        $Q_1 R_1$ & $Q_1 M_r W_1$ \\
		$Q_2 R_0$ & $Q_x M_x W_x$ \\
		$Q_2 R_1$ & $Q_3 M_r W_0$ \\
		$Q_3 R_0$ & $Q_3 M_x W_0$ \\
		$Q_3 R_1$ & $Q_3 M_x W_1$ \\
        \hline
    \end{tabular}
    \label{tab:tmadd}
\end{table}

Thus, presented with a number of the form $1^l01^m$ on the tape from left to right at position $0$ with the initial state of $Q_0$, it returns $1^{l+m}$ and stays in the static (halting, accepting) state of $Q_3$.
The $x$ in $M_x$ for $Q_3$ state indicates don't care.
The state $Q_2R_0$ is unreachable for this formulation.
$Q_n$ takes 2-bits while $M$ and $W$ takes 1-bit each, thus, there are a total of $2^{(4*2)*(2+1+1)} = 4294967296$ possible programs of the same length.

The Universal Turing machine modifies the Turing machine to drop the requirement to have a different transition table for each application.
The program/table can also be read as input from the tape itself like the von-Neumann stored program concept.
This employs a universal computer model that can do all physical computation as defined by the Church-Turing thesis.

The initial model of quantum computation extended this by the Church-Turing-Deutsch thesis, which defined the Quantum Turing Machine.
This allows representing the tape in a superposition of basis states.
The computation, i.e. the transition table, is defined as a unitary function that maps onto the basis state.
This model is equivalent to the more general circuit model of computation.
An example of this is the quantum adder, which can evolve in superposition two inputs and produce the result as a superposition till it is collapsed by measurement.

\begin{figure}[hb]
    \centering
    % trim: LBRT
    \includegraphics[clip, trim=14cm 0cm 0cm 8cm, width=1.0\columnwidth]{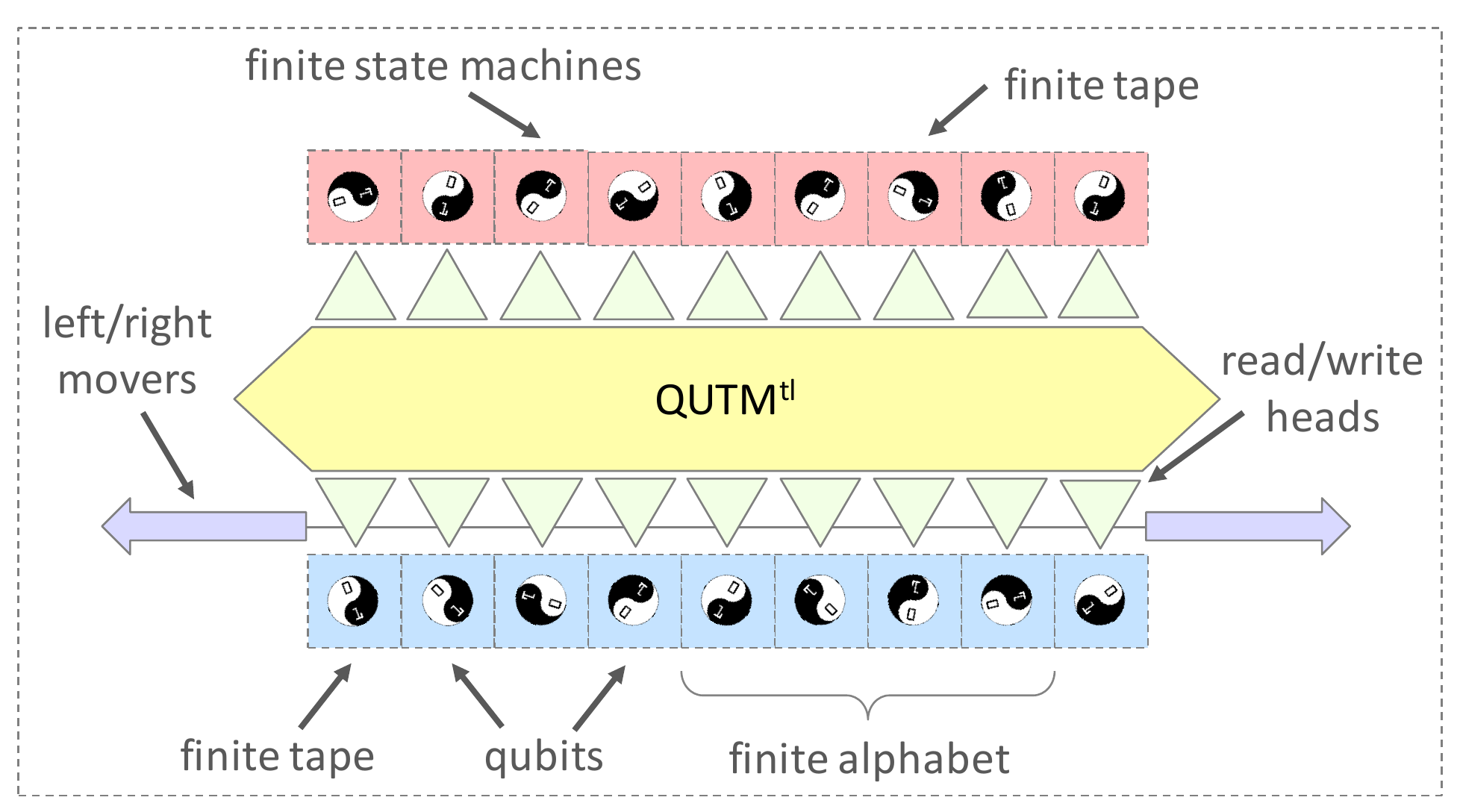}
    \caption{Our computation model: Quantum Universal Turing Machine time-length restricted}
    \label{fig:qutm-tl}
\end{figure}

We further extended this QTM model to circumvent the requirement of a fixed unitary function for each program, by modelling the UTM on quantum circuit.
We call this model the $\text{QUTM}^{\text{tl}}$, a time and length restricted quantum universal Turing machine, as shown in Fig.~\ref{fig:qutm-tl}.
It can be understood as having the same relation to $\text{QTM}^{\text{tl}}$ as a programmable-computer has to an embedded system.

\begin{figure}
    \centering
    % trim: LBRT
    \includegraphics[clip, trim=47cm 0cm 0cm 54cm, width=1.0\columnwidth]{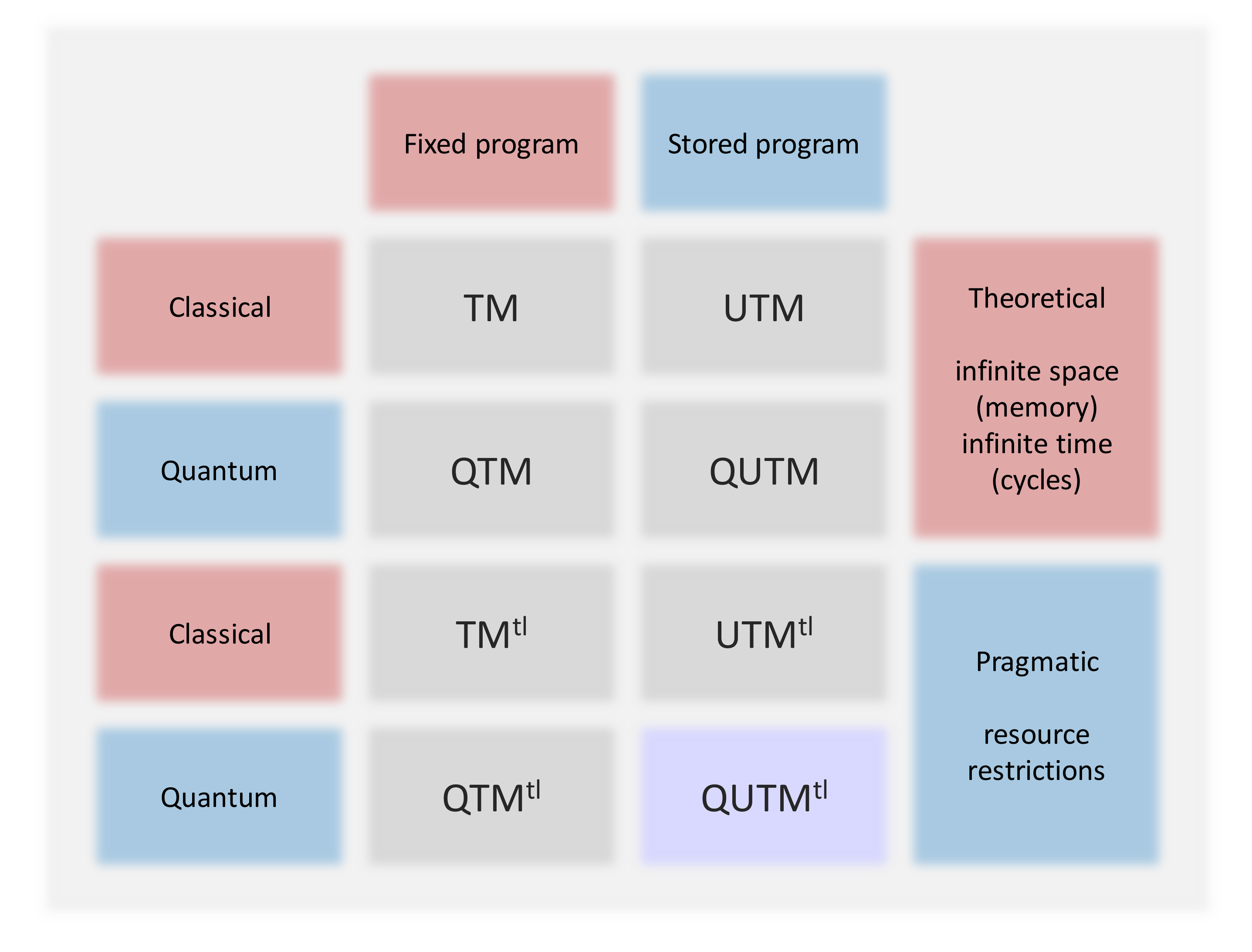}
    \caption{Variants of Turing Machine (TM) with respect to (i) classical or quantum, (ii) program hard-coded or reprogrammable by storing, and (iii) theoretical or pragmatic model}
    \label{fig:tmvar}
\end{figure}

Each iteration of the TM undergoes the following transforms:
\begin{enumerate}[nolistsep,noitemsep]
    \item $\{q_{read}\} \leftarrow U_{read} (\{q_{head}, q_{tape}\})$
	\item $\{q_{write}, q_{ch}, q_{move}\} \leftarrow U_{\delta} (\{q_{read}, q_{state}, q_{\delta}\})$
	\item $\{q_{tape}\} \leftarrow U_{write} (\{q_{head}, q_{write}\})$
	\item $\{q_{head}\} \leftarrow U_{move} (\{q_{head}, q_{move}\})$
\end{enumerate}
Our model~\cite{sarkar2020quantum} allows to evolve a superposition of programs on an entangled superposition of inputs.
Thereby it completes the automata variants with respect to the three factors as shown in Fig.~\ref{fig:tmvar}.

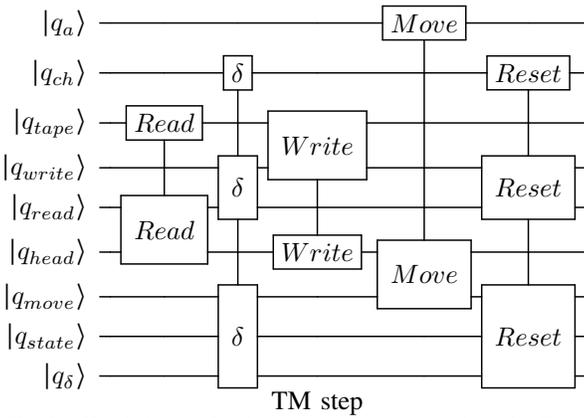
\begin{figure}[htb]
% \centering
\raggedright
\centerline{
\Qcircuit @C=0.4em @R=0.6em {
\lstick{\ket{q_{a}}}	 	& \qw 	& \qw 							& \qw 								& \qw 					& \gate{Move} \qwx[5] 	& \qw 							& \qw \\
\lstick{\ket{q_{ch}}} 		& \qw 	& \qw 							& \gate{\delta} \qwx[2] 			& \qw 					& \qw 					& \gate{Reset} \qwx[2] 			& \qw \\
\lstick{\ket{q_{tape}}} 	& \qw 	& \gate{Read} 					& \qw 								& \multigate{1}{Write} 	& \qw 					& \qw 							& \qw \\
\lstick{\ket{q_{write}}} 	& \qw 	& \qw 							& \multigate{1}{\delta} 			& \ghost{Write} 		& \qw 					& \multigate{1}{Reset} 			& \qw \\
\lstick{\ket{q_{read}}} 	& \qw 	& \multigate{1}{Read} \qwx[-2] 	& \ghost{\delta} 					& \qw 					& \qw 					& \ghost{Reset} 				& \qw \\
\lstick{\ket{q_{head}}} 	& \qw 	& \ghost{Read} 					& \qw 								& \gate{Write} \qwx[-2]	& \multigate{1}{Move} 	& \qw 							& \qw \\
\lstick{\ket{q_{move}}} 	& \qw 	& \qw 							& \multigate{2}{\delta} \qwx[-2] 	& \qw 					& \ghost{Move} 			& \multigate{2}{Reset} \qwx[-2]	& \qw \\
\lstick{\ket{q_{state}}} 	& \qw 	& \qw 							& \ghost{\delta} 					& \qw 					& \qw 					& \ghost{Reset} 				& \qw \\
\lstick{\ket{q_{\delta}}} 	& \qw 	& \qw 							& \ghost{\delta} 					& \qw 					& \qw 					& \ghost{Reset} 				& \qw \\
& & & & \mbox{TM step} \\
}
}
\caption{Circuit blocks for the quantum implementation of a TM step in the stored-program model}
\label{qcirc:qutm}
\end{figure}

The quantum circuit primitive for $\text{QUTM}^{\text{tl}}$, as shown in Fig.~\ref{qcirc:qutm} is employed for estimating algorithmic metrics which involves evaluating the statistical behaviour of a enumerable set of programs.

\subsection{ACSS-q}

The model from \cite{sarkar2020quantum} is modified to consider the CTM application.
The modifications include adding a halting state, chosing the initial state, defining a Turing tape length and setting the number of iterations.

A Halting state $Q_h$ is needed for calculating the CTM.
$Q_0$ is chosen as the Halting state.
For the Halting state, the state loops on itself, overwrites the read character with itself and moves arbitrarily to left or right.
$$Q_h \times R_r \rightarrow Q_h \times W_r \times M_l$$

The initial state $Q_{init}$ also needs to chosen to be different from the Halting state (unless it is for the trivial case of $m=1$ states).
It can be set to the all-1 state by applying an X-gate on all the current-state qubits.

% \colorbox{green}{How many accepting states Zenil considers?}

The size of the Turing tape $c$ is set based on the length of strings we want to calculate the algorithmic complexity of, instead of being the same as the description length\cite{sarkar2020quantum} (for detecting self-replicating machines).
This is based on the block size extension we want to do for BDM.

While the number of iterations $z$ should be same as the Busy Beaver run length, Calude showed that we miss exponentially fewer machines for smaller run length.
Since the CTM is based on the ratio of halting machines to accepting machines, it can be well approximated with much lower runtimes.
The number is related to the block size as we want most strings to be represented in the output.
It also depends on the number of states, as a small number of states and a large block, might not well represent the universal distribution even for large iterations.
A formal treatment of setting these hyper-parameters needs to be considered as further research.

Thus, for the chosen value of $m$ all $2$-state $1$-dimensional Turing machines can be run in superposition.
After $z$ steps, the measurement of the state vector estimates the probability $P_h$ of the current state register to hold $Q_h$; and the probability $P_s$ of the $c$ sized tape to hold $s$ among them.
The ratio of $P_s/P_h$ is the estimate for the CTM calculation of $s$.

\begin{sidewaysfigure*}
    \centering
    \includegraphics[clip, trim=0cm 0cm 0cm 3cm, width=1.0\textwidth]{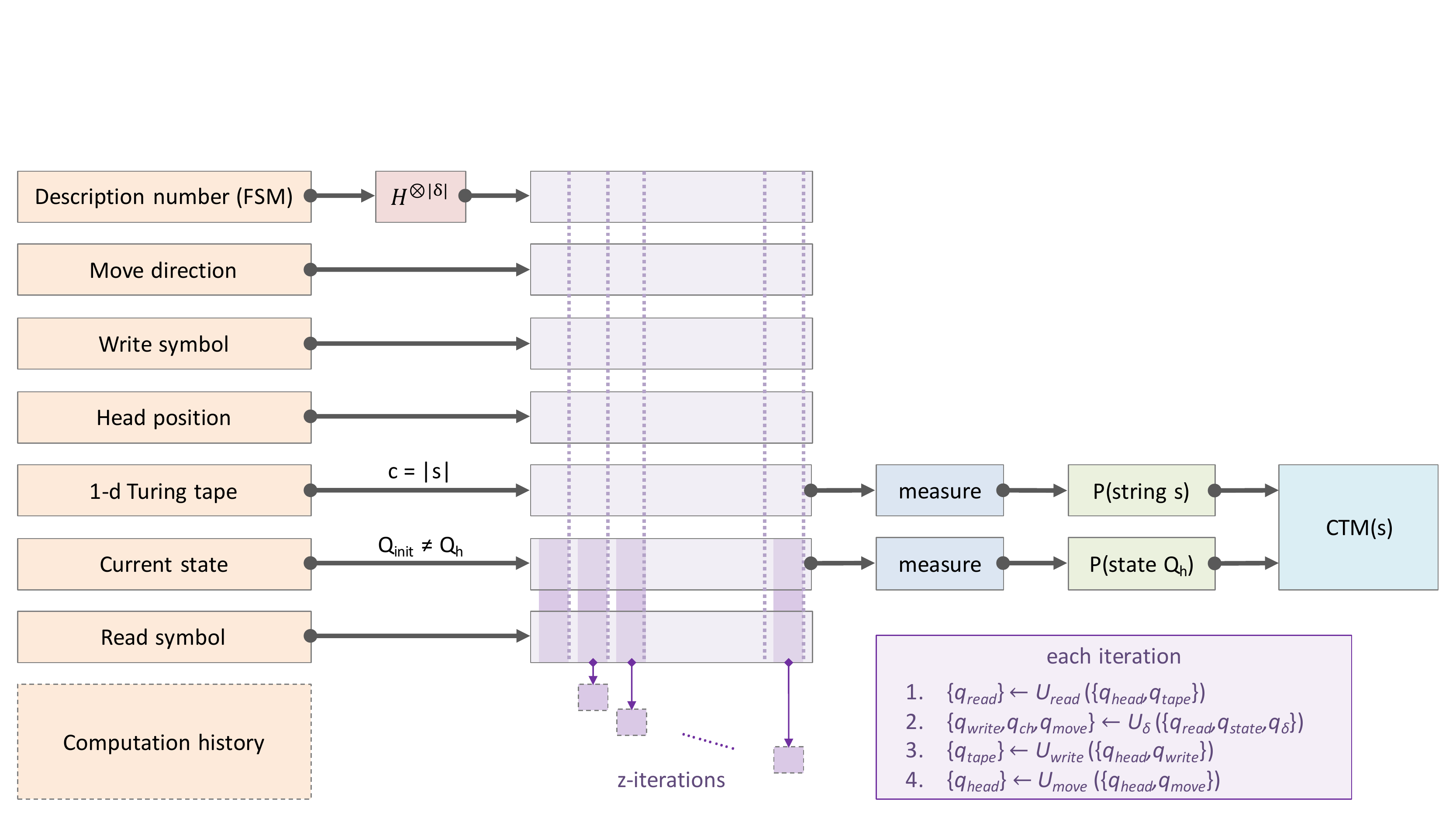}
    \caption{ACSS-q framework}
    \label{fig:framework}
\end{sidewaysfigure*}

% \begin{figure}[ht]
%     \centering
%     \captionsetup{justification=centering}
%     % trim: LBRT
%     \includegraphics[clip, trim=0cm 0cm 0cm 3cm, width=1.0\textwidth]{figures/acssq1.pdf}
%     \caption{ACSS-q framework}
%     \label{fig:framework}
% \end{figure}

\subsection{Quantum resource estimate}

The qubit complexity of the implementation (assuming $q_a$ ancilla qubits) is:
\begin{align*} 
q_{QUTM} &=  q_\delta+q_{state}+q_{move}+q_{head}+q_{read}+q_{write}\\&+q_{tape}+q_{ch}+q_a \\ 
        &= (m*n*(log(m)+log(n)+d))+log(m)\\&+d+log(c)+log(n)+log(n)+(c*log(n))\\& +((z-1)*(log(m)+log(n)))+q_a
\end{align*}
Where the log is taken base-2 and rounded up to the nearest integer.

Let us consider 3 cases here to calculate the quantum resource needed:
\begin{itemize}[nolistsep,noitemsep]
    \item For $d=1$, $n=2$, $m=5$ states, $c=12$ length strings
        \begin{align*} 
        q_{QUTM} &=  (5*2*(log(5)+log(2)+1))+log(5)\\&+1+log(12)+log(2)+log(2)\\&+(12*log(2))\\& +((z-1)*(log(5)+log(2)))+q_a \\
                % &=  (5*2*5)+3+1+4+1+1+(12*1)+((z-1)*4)+q_a \\
                &=  72+((z-1)*4)+q_a
        \end{align*}
    \item For $d=1$, $n=2$, $m=5$ states, $c=13$ length strings
        \begin{align*} 
        q_{QUTM} &=  (5*2*(log(5)+log(2)+1))+log(5)\\&+1+log(13)+log(2)+log(2)\\&+(13*log(2))\\& +((z-1)*(log(5)+log(2)))+q_a \\
                % &=  (5*2*5)+3+1+4+1+1+(13*1)+((z-1)*4)+q_a \\
                &=  73+((z-1)*4)+q_a
        \end{align*}
    \item For $d=1$, $n=2$, $m=6$ states, $c=12$ length strings
        \begin{align*} 
        q_{QUTM} &=  (6*2*(log(6)+log(2)+1))+log(6)\\&+1+log(12)+log(2)+log(2)\\&+(12*log(2))\\& +((z-1)*(log(6)+log(2)))+q_a \\
                % &=  (6*2*5)+3+1+4+1+1+(12*1)+((z-1)*4)+q_a \\
                &=  82+((z-1)*4)+q_a
        \end{align*}
\end{itemize}

For real-world applicability, the logical qubits in $\approx10^2$ order of magnitude is quite low (compare Shor's algorithm for RSA-512).
Most importantly, we find that, for linear change in state and length, the number of qubits grow linearly/linearithmically.
This is in huge contrast with the super-exponential complexity in classical simulation which prohibits extending the ACSS beyond the $5$-state $12$-length bound.

Infact, the biggest contribution to the qubit complexity is the growing computation history log adding $n+log_2(m)$ qubits to the superposition.
To execute $500$ iterations in the first case, this would take $2000$ qubits.
This is unavoidable to maintain the reversibility~\cite{benioff1982quantum} and various tricks in reversible compilation~\cite{meuli2019reversible} uses techniques like repeat-until-success, dirty ancilla, etc. for space-time trade-offs.
Since the computation history is not useful except to maintain the reversibility, this can be circumvented by tracing them out and restarting the computation after few steps.
Given there are space-constraints in the number of qubits allowing us to store the computation history only upto $z_{ch} < z$, the entire run of the algorithm can be divided into stages.
In the first run, the computation is carried out for $z_{ch}$ steps and the final state for head, tape and current state is read out.
The computation can be restarted with a blank computation history by preparing this state for the corresponding registers (head, tape, current state) and continuing the $z_{ch}+1$ iteration.

% There is a different method, where we swap reset the tape write symbol at every step to a new "tape update log" and uncompute the read and current state immediately. This will require only z qubits.
% After the swap all-zero state to write, then uncomputer tape head, then uncompute tape 

It is needless to mention that this application is not targeted towards noisy intermediate-scale quantum (NISQ) computing.
The logical qubits we consider in our analysis are perfect (i.e. does not decohere) and in a fully-connected (complete graph) topology.
The quantum circuit can be tested as a proof-of-concept with quantum computing simulators like QX.
These simulators allow introducing chip topology and errors models towards a more realistic simulation for NISQ.
Thus, realistic implementations would also consider qubit mapping, routing and error-correction features built in the quantum compiler like OpenQL.
However, considering these abstractions would make this application intractable on QX and current NISQ hardware; and the application design presented in this work remain agnostic to these hardware layer limitations. % Quantum implementation
% \newpage
\section{Applications} \label{sec:applications}

Theoretical applications of Kolmogorov complexity is widespread~\cite{li2008introduction}, finding various uses in artificial general intelligence, theoretical physics, psycology, data compression, finance, linguistics, neuropsychology, psychiatry, genetics, sociology, behavioral sciences, image processing, among others.
However, estimating algorithmic information for practical datasets is often computationally intractable due to the large number of enumerations (as shown in Table~\ref{tab:numP}) that needs to be executed.
The exploration in this research is towards to field of Experimental Algorithmic Information Theory (EAIT)~\cite{zenil_2020} which is growing in popularity due to its ubiquity in tasks that can be modeled as inductive reasoning.
It connects theoretical computer science to the real-world by a quasi-empirical approach to mathematics (popularized as meta-mathematics by Gregory Chaitin, one of the founders of AIT).
It involves enumerating and running programs to understand its statistical mechanics and is similar to Stephen Wolfram's Computational Universe approach~\cite{wolfram2002new}.
Here we present two potential application of the ACSS-q framework in the computational bioinformatics domain, a near-term and a prospective usecase.

\subsection{SARS-CoV-2 and human PPI}

There has been many successes of EAIT in recent years, especially in the field of biological sequence analysis.
These techniques expand our understanding of the mechanisms underlying natural and artificial systems to offer new insights.
Algorithmic Information Dynamics is at the intersection of computability, algorithmic information, dynamic systems and algebraic graph theory to tackle some of the challenges of causation from a model-driven mechanistic perspective, in particular, in application to behavioral, evolutionary and molecular reprogramming.

A recent exploration towards this is in understanding the protein-protein interaction (PPI) map~\cite{gordon2020sars} between the SARS-CoV-2 proteins in human cells (the coronavirus responsible for the Covid-19 pandemic) and human proteins.
332 high-confidence PPI were experimentally identified using affinity-purification mass spectrometry.
However, the mechanistic cause behind these specific interactions is not a well understood phenomena yet.
A recent work\footnote[3]{Slide 51, presented by Dr. Alyssa Adams in AUTOMATA 2020, yet to be published~\cite{adams_2020}} tries to explore the Kolmogorov complexity estimates of these PPI and found a positive correlation in the BDM values of the interactions.
Such studies will help us predict in-silico the biological dynamics, helping us find drug targets for therapeutics.

The BDM used for the study is based on the ACSS database, limited to the block length of 13.
Extending ACSS to larger block lengths with help bridge the causal gap, which, for longer strings like proteins can be considerable, limiting its advantage over traditional entropy based methods.
The quantum algorithm described in this paper can potentially extend the ACSS database to empower the BDM more towards the actual CTM value.

\subsection{In-quanto synthetic biology}

The ability to design new protein or DNA sequences with desired properties would revolutionize drug discovery, healthcare, and agriculture. 
However, this is challenging as the space of sequences is exponentially large and evaluating the fitness of proposed sequences requires costly wet-lab experiments.
Ensemble approaches~\cite{angermueller2020population} with various machine learning (ML) methods, mostly generative adversarial networks (GAN), suitable to guide sequence design are employed in-silico, instead of relying on wet-lab processes during algorithmic development.
Since the mechanism for determining the fitness of the model is known, it can be encoded as a quantum kernel that evaluated in superposition the population of sequences.
This is part of our future exploration in using the framework developed in this paper for in-silico (or, in-quanto) synthetic biology.

The fitness is often based on Sequence-based Genome-Wide-Association-Studies (GWAS) that combines sequence data to expressed phenotypes.
These databases are increasingly growing with the advent of low-cost sequencing tools, however, analysis of this huge data volume for inference and application remains a challenge.
Gene Ontology (GO) curates the knowledge of species-agnostic gene products with respect to molecular function, cellular component and biological process annotations.
% It is designed to be includes prokaryotes, eukaryotes, single and multicellular organisms.
% Molecular function terms describe activities that occur at the molecular level, e.g. catalysis and transport.
% Cellular component annotates the locations relative to cellular structures in which a gene product performs a function, either cellular compartments (e.g., mitochondrion), or stable macromolecular complexes of which they are parts (e.g., the ribosome).
% Biological process refers to the larger processes in the biological programs accomplished by multiple molecular activities, e.g. DNA repair or signal transduction.
% An example of GO annotation for the gene product "cytochrome c" can be described by the molecular function of oxidoreductase activity, the biological process of oxidative phosphorylation and the cellular component as mitochondrial matrix.
Similarly, Sequence Ontology~\cite{eilbeck2005sequence} (SO) is a set of terms and relationships used to describe the features and attributes which can be located on a biological sequence.
It is designed to make the naming of DNA sequence features and variants consistent and therefore machine-processing.
They are defined by their disposition to be involved in a biological process, e.g. binding site and exon.
It forms a structured controlled vocabulary for the description of primary annotations of nucleic acid sequence.

SO, sequence-based GWAS, GAN needs to be combined in-quanto towards future synthetic biology techniques. % Applications
% \newpage
\section{Conclusion} \label{sec:conclusion}

In this paper we present the ACSS-q framework, a quantum accelerated method to extend the algorithmic complexity of short string database.
It uses the Coding Theorem Method (CTM) to calculate the Kolmogorov complexity using the Solomonoff algorithmic probability of the string being generated by a random program a universal Turing machine.
This database provides an estimate of the description complexity of longer strings that can be cumulated by the Block Decomposition Method (BDM) for comparative analysis using the ACSS value of shorter strings.
These methods can bridge the causal gaps in data with respect to other methods based on entropy and compression.
The complexity value reflect the natural occurrence of these strings and useful for building generative models.

The number of program enumerations required for CTM grows super-exponentially, quickly becoming computationally intractable on classical supercomputers even for strings of size 13, limiting the discovery of causation for longer strings.
A quantum circuit formulation for simulating a superposition of Turing machines to estimate the algorithmic probability is explored in this research.
The number of qubits is proportional to the description length and grow linearly with the length of the short string.
This work builds upon various results in estimating algorithmic metrics with resource bounds of memory and cycles.

We present a use-cases where the state-of-the-art classical result for SARS-CoV-2 and human the protein-protein interaction can be improved by our quantum technique leading to applications in drug design and also for synthetic biology in the future.
This algorithm can be implemented gradually on increasingly larger quantum computing platforms based on the maturity of the field, allowing the refinement of experimental algorithmic information theory results to converge to their descriptive complexity. % Conclusion

% \appendix
% \newpage
% \appendixpage
% \input{appendix_1} % Motivation

% \newpage
\bibliographystyle{unsrt}
\bibliography{references/ref.bib}

\end{document}